\pdfoutput=1
\documentclass[letterpaper, reqno]{amsart} 
\usepackage{epsfig}
\newtheorem*{ack}{Acknowledgment}

\newcommand{\ket}[1]{| #1 \rangle}

\newcommand{\pic}[5]{\raisebox{#3pt}{\hspace{#4pt} \epsfig{file=#1.pdf,height=#2pt,silent=} \hspace{#5pt}}}

\newcommand{\kd}[1]{\mathchoice{
\pic{#1}{24}{-8}{-1}{2}}{
\pic{#1}{11}{-3}{1}{1}}{
\pic{#1}{9}{-2}{-3}{1}}{
\pic{#1}{7}{-1}{-1}{0}}}

\newcommand{\lkd}[1]{\mathchoice{
\pic{#1}{55}{-25}{-1}{2}}{
\pic{#1}{25}{-11}{1}{1}}{
\pic{#1}{20}{-10}{-3}{1}}{
\pic{#1}{18}{-8}{-1}{0}}}
\begin{document}
 
\title{On the q-quantum gravity loop algebra} 
\author{Seth Major} 
\date{January 2008} 
\address{Department of Physics\\ 
Hamilton College\\ 
Clinton NY 13323 USA} 
\email{smajor@hamilton.edu} 

\begin{abstract} 
A class of  deformations of the $q$-quantum gravity loop algebra is shown to be incompatible with the combinatorics of Temperley-Lieb recoupling theory at a root of unity.  This incompatibility appears to extend to more general deformation parameters.
\end{abstract} 
\maketitle

In the Ashtekar formulation of complex self-dual connections and conjugate densitized triads, general relativity takes on a particularly simple form; the constraints are polynomial in the elementary variables. With a particular choice of operator ordering, the Kodama state satisfies these constraints \cite{kodama}. Built from the Chern-Simons functional $S_{cs}[A] = \int_\Sigma \text{Tr}[  A \wedge dA + \tfrac{2}{3} A \wedge A \wedge A ]$ the Kodama state is
\[
\psi[A] = { N} \exp \left( \frac{3}{\ell_p^2 \Lambda} S_{cs} [A] \right)
\]
in which ${ N}$ is a - possibly topology-dependent - normalization constant \cite{soo} and $\Lambda$ is the cosmological constant. This state has de Sitter spacetime as a semi-classical limit and so it provides a possible state for quantum gravity with positive cosmological constant.  See \cite{leerev} for an extensive review.

Since the usual basis for states in loop quantum gravity are spin networks, it is natural to define the loop (or spin network) transform of the Kodama state
\begin{equation}
\label{looptrans}
I=\int d\mu[A] \psi[A] \phi_s[A] = { N} \int d\mu[A] \phi_s[A] \exp \left( \frac{3}{\ell_p^2 \Lambda} S_{cs} [A] \right)
\end{equation}
for a spin network $\phi_s[A]$. As for any integral transform, the appropriate definition of the integral requires some care.

In the Euclidean context the transform involves the delightful convergence of gravity, topological quantum field theory, and knot theory.  The definition of the transform was provided, in part, by Witten who found that the transform of links was equal to the Kauffman bracket \cite{witten}. He found that the integral must be regulated; the simple one dimensional excitations of geometry must be ``thickened" into ribbons, or framed loops.  This suggests that the spin network representation of the Kodama state requires  framed spin-networks.  

As originally formulated \cite{qqg, qqgop, dis}, $q$-quantum gravity is a modification of the original loop representation of loop quantum gravity (LQG) that takes this framing into account. The framing dependence is encoded in functions of the complex phase $q=e^{i \pi / r}$, which depends on the cosmological constant; the parameter $r$ is inversely related to the dimensionless cosmological constant $r \propto 1/\Lambda \, \ell_p^2 $.  $Q$-quantum gravity is based on the conjecture that the loop representation of the Kodama state - quantum gravity with a cosmological constant - has a quantum spin network basis. The recoupling theory of these ``$q$-spin nets" is, at least for simple, planar vertices, given by Temperley-Lieb recoupling theory as described in Kauffman and Lins \cite{KL} (See the appendix for a summary.)   

The picture, if it could be completed, is a remarkable one:  The states of quantum gravity with a cosmological constant would be described by knot invariants expressed in terms of $q$-spin networks.  However, several problems impede progress in completing the picture: (i)  We lack a definition - even for links - of the loop transform for complex connections required of the Lorentzian theory. 
(ii) In the linearized Lorentzian theory the state is non-normalizable \cite{LeeLaurent}.  
(iii) At least in the context of variational calculus, the loop transform picks up a sensitivity to the tangent space structure at vertices \cite{vertex}. 
(iv) The Chern-Simons form of the Kodama state is not CPT invariant. And
(v) The Kodama state is clearly not a state in usual framework of LQG. For some detailed comments on the nature of the difficulties see the appendix of \cite{thomasrev}. 

As it stands $q$-quantum gravity is a kinematic theory.  Even at the kinematic level, one might wonder whether there exists a unique formulation of the loop algebra.  Ten years ago the following question was posed, Is there a deformation of the loop algebra consistent with Temperley-Lieb recoupling theory? \cite{dis}.  I show in this note that the answer is essentially no.\footnote{The  ``essentially" refers to the fact that the $r=5$ case is consistent and, to arbitrary accuracy, the algebra is satisfied for small, integer cosmological constant and representations $n\ll r$.}  In fact the argument presented here shows that the basic loop algebra is inconsistent with Temperley-Lieb recoupling theory (at a root unity). 

In the following I briefly review the loop representation of $q$-quantum gravity and then present the combinatorial argument.  All the representations are labeled with integers, rather than the $1/2$-integers of angular momentum.
I use the term $q$-classical to indicate the classical limit of the $q$ deformation for which $q=1$ when either (or both) $\hbar G \rightarrow 0$ or $\Lambda \rightarrow 0$.

In $q$-quantum gravity \cite{qqg}, the basic operators are directly defined in terms of their action on a $q$-spin network.  The $SU(2)_q$ loop operator $T_q[\beta]$, based on the framed loop $\beta$, acts by raising and lowering the representation on the edge; if a loop $\beta$ carries a representation $n$, denoted here by $\ket{n}$, then
\begin{equation}
T_q[\beta] \ket{n} := \ket{n+1} + \ket{n-1}.
\end{equation}
For a cycle $\beta$ in a network, the action of $T_q[\beta]$ is found using recoupling theory.  $T^a_q[\alpha]$, the ``momentum" operator based  the framed loop $\alpha$, may be defined by the action on an arbitrary $q$-spin network \cite{dis}.  For the argument discussed here it is sufficient that the network $\Gamma$ does not contain the loop $\alpha$ and that $\alpha$ intersects an edge in $\Gamma$ once. Then 
\begin{equation}
\label{t1}
T_q^a[\alpha](s) \ket{\Gamma} := - i \,n \, \ell_p^2 \Delta^a \left[ \Gamma,
\alpha \right](s) \, \ket{ \Gamma  \alpha },
\end{equation}
when the edge carries representation $n$. The distribution 
\[
\Delta^a [\Gamma , \alpha ] (s) :=  \int dt \, \delta^3 
\left[ e (t) , \alpha (s) \right] \dot{e}^a (t)
\]
is inherited from the classical loop algebra.\footnote{This is the Euclidean version of the operator.}  The notation $\Gamma \alpha$ indicates that the spin network was enlarged to include loop $\alpha$ and a new 4-valent intersection - the operator ``grasps" the $q$-spin network.\footnote{The intertwiner of this vertex is given by a edge, labeled with 2, connecting the loop $\alpha$ with the loop $\beta$. See \cite{carlobook} page 250 for a more detailed description in the usual LQG context.}  This definition may be easily extended to an arbitrary state \cite{qqg,dis}.  

The operators should satisfy two requirements: (1) The usual algebra is recovered in the $q$-classical limit; and (2) The elementary algebra is closed. The first requirement is obviously satisfied, given the definitions and the fact that $SU(2)$ spin network recoupling is recovered in the $q$-classical limit.  However, other choices for the action of $T^a_q[\alpha]$ also satisfy this requirement.  For instance replacing $n$ in equation (\ref{t1}), which represents the result of grasping an edge with label $n$, with the quantum integer $[n]$ also satisfies (1).  This note asks,  Does there exist any action or $T^a_q[\alpha]$, whether involving $n$, $[n]$, or any other factor,  that is consistent with both (2) and Temperley-Lieb recoupling theory?     

With these definitions for the loop operators, the elementary loop algebra given in the original formulation \cite{qqg, dis} is
\[
\left[ T^a_q[\alpha](s), T_q[\beta](s) \right] =  - i \ell_p^2  \,  
\Delta^a[\alpha,s] \, T_q[\alpha  \beta].
\]
for loops $\alpha$ and $\beta$ that intersecting once.  As above ``$\alpha  \beta$" denotes the $q$-spin network constructed from the two loops.  For further detail, see \cite{qqg, dis}. 

To make the argument more general, let us consider the $q$-deformed algebra or ``qummutator'' $[a,b]_\lambda := ab
-\lambda ba$.  The general loop algebra is
\begin{equation} 
\label{qumator}
\left[ T^a_q[\alpha](s), T_q[\beta](s) \right]_\lambda =  - i \eta \, 
\ell_p^2 \Delta^a[\alpha,s] T_q[\alpha \beta].
\end{equation}
where both $\eta$ and $\lambda$ are free parameters that equal 1 in the $q$-classical limit.  (A further generalization in which $\lambda$ is also an operator is discussed at the end of this note.) As is easy to check, the parameter $\eta$ simply re-scales the possible solutions in an identical manner as the initial value in a series solution.   Hence $\eta$ is set to 1 in the following.

The question mentioned above now becomes: Is Temperley-Lieb combinatorics consistent with the generalization of the $T$-algebra given in equation (\ref{qumator})?
Since this is question of combinatorics, I suppress the unnecessary constants and distributional factors.  For the purposes of this argument it is sufficient to define operators $T$, $S_\alpha$, and $S_{\alpha \beta}$ that act on a state $\ket{ n, \, 0}$, a function of the two intersecting loops, $\alpha$ and $\beta$ with representations $n$ and $0$, respectively. (The definition for an arbitrary state is given in the appendix, equations (\ref{tdef}) and (\ref{sdef}).)  
\begin{equation}
\begin{split}
T  \ket{ n, \, 0} & := t_n \ket{ n, \, 1}\\
S_\alpha \ket{n, \, 0} &:= \ket{n+1, \, 0} + \ket{n-1, \, 0}
\end{split} 
\end{equation}
For notational simplicity I substituted the operator $S_{\alpha}$ for the $T_q[\alpha]$ operator and will denote $S_{\alpha \beta}$ for $T_q[\alpha \beta]$. The $T$ operator is the combinatorial portion of the $T_q^a[\beta]$ operator of equation (\ref{t1}), i.e. the bit that gives $n$ in the original definition.

The argument is based on recursion relations for the $t_n$'s in terms of quantum integers and the parameter $\lambda$.  These relations may be generated by calculating the action of the deformed algebra of (\ref{qumator}) on the state $\ket{n, \, 0}$.  The deformed algebra in terms of the operators $T$ and $S_\alpha$ is
\begin{equation}
\label{alg}
\left[ T, S_\alpha \right]_\lambda \ket{n, \, 0} = S_{\alpha \beta} \ket{n, \, 0}.
\end{equation}
Making use of the graphical methods of Temperley-Lieb recoupling theory the calculation is straightforward. One way is described in the appendix, equations (\ref{algreduc1} - {\ref{algreduc2}). The loop algebra reduces to
\begin{equation}
\begin{split}
t_{n+1} \ket{ n+1, \, 1} + t_{n-1} \ket{n-1, \, 1} - \lambda t_n \frac{[n+2][n-1]}{[n][n+1]}  \ket{n-1, \, 1} - \lambda t_n \ket{ n+1, \, 1}  \\
=  \ket{n+1, \, 1} - \frac{[n-1]}{[n+1]} \ket{n-1, \, 1}
\end{split}
\end{equation}
which immediately yields the pair of recursion relations
\begin{align}
\label{recursion}
&t_{n+1} - \lambda t_n = 1 , &\text{for  } 1 \leq n \leq r-3 \\
\label{recursion2}
&t_{n-1} - \lambda  \frac{[n+2][n-1]}{[n][n+1]} t_n = - \frac{[n-1]}{[n+1]}, &\text{for  } 2 \leq n \leq r-2
\end{align} 
The upper limit on $n$ is due to the limit on the representations at $r-1$ and the result that $[r]=0$.  When $n=r-2$ the second equation contains the result that 
\begin{equation}
\label{tlimit}
t_{r-3} = -\frac{[r-3]}{[r-1]}
\end{equation}
The first equation immediately gives $t_n = 1+\lambda + \lambda^2 + ... + \lambda^{n-1} = (1- \lambda^{n})/(1-\lambda)$ -- a ``$\lambda$-deformed" integer, $\{ n \}_\lambda$. It is also clear that the recursion relations are satisfied for the $q$-classical limit with $\lambda=1$. 

It is clear that a deformation is possible only if the two recursion relations are consistent. Using the result for $t_n$ from the first equation in the second, the problem then reduces to simply whether a solution (or solutions) for $\lambda$ exist.  By solutions I mean whether a value for $\lambda$ exists for any integer $r$ and all $2 \leq n \leq r-2$. (This is were the argument replies on $q$ being (primitive) root of unity.) Starting from small $r$ we see that the cases $r<4$ are trivial.  The case $r=4$ has no solutions since $\lambda$ doesn't appear and $t_1 = 1$ and, from the second equation, $t_1 = -1$. For $r=5$, $\lambda = -2.618...$ is a solution for the two polynomials for $n=2$ and $n=3$ generated by the second equation.  For $r=6$ we see from the solutions for $\lambda$ :
\begin{align*}
n&=2  &\lambda &= ( -2.30278, 1.30278 ) \\
n&=3  &\lambda &= (-1.24279 - 1.07145 i,   -1.24279 + 1.07145 i, 1.48558 ) \\
n&=4  &\lambda &= (-0.5 - 1.65831 i,  -0.5 + 1.65831 i)
\end{align*}
that no solution exists.  Further investigation shows that this trend continues with an exception.  For large $r$, there is an approximate solution of $\lambda =1$.  This is expected in the $q$-classical limit, as $r \rightarrow \infty$  
\[
[n] \approx n \left(1 - \frac{1}{6} \left( \frac{\pi}{r} \right)^2 \left( n^2 -1 \right) \right).
\]    
For large $r$, and not so large $n$, the approximate solution is given by $\lambda \sim 1$.  For instance, for $r=80$ and the representations $n$ from 2 to 38, the roots of the polynomial differ from 1 by a part in $10^3$.  If $r \sim 10^5$, as during inflation, similar roots for $n=2$ and $n=100$ differ by a part in $10^{14}$. We see that consistent solutions only occur for $r=5$ and for $r \rightarrow \infty$, when the classical loop algebra is recovered for $\lambda =1$. 

This argument may be extended in several ways. (i) If the transform for the Lorentzian theory requires analytic continuation of $r$, the result is similar.  There are apparently no solutions for complex $r$. The approximate $\lambda=1$ solution remains for small cosmological constant.
(ii)
Although the argument is expressed in terms of the loop algebra it is easy to use extend the argument to the flux-holonomy  algebra as reviewed in \cite{thomasrev}.  In fact the action is quite a bit more simple in these variables. 
(iii)
A more general deformed algebra could contain an operator $\lambda$.  If $q$-spin networks are also eigenvectors of this new operator then there are two cases to consider. The new operator could act to the right or left of the $T_q[\alpha]$ operator.  The resulting eigenvalues are functions $\lambda(n)$ or $\lambda(n \pm 1)$, respectively.  In either case the recursion relations are simple generalizations of equations (\ref{recursion}) and (\ref{recursion2}), now with a $n$-dependent $\lambda$.  The resulting non-linear recursion relations may be solved.  However, they depend on 2 or 3 arbitrary initial values, respectively, in the two different choices of operator ordering.  So there is no unique solution.  In addition the relation between these solutions and the $SU(2)$ loop algebra is not clear.  This is also the case for more general operator-based deformations such as those based on loops; requirement (1) apparently rules these out.
(iv)
One may be tempted to take the result $t_n = \{n\}_\lambda$ seriously and ask, under what conditions is $t_n$ simply related to $[n]$, as would be natural for an eigenvalue of a $q$ deformed operator.  A short calculation using equations (\ref{recursion}) and (\ref{recursion2}) shows that $\lambda=q^2$ and $t_n = [n] q^{n-1}$.  Further, by multiplying the right hand side of the second recursion relation by -1 and the first term by $q^4$ , the system is consistent for all integer $r$.  Since these factors change the loop algebra, this does not provide a viable solution to the question.

To conclude, the above argument shows that there does not exist a $T$-algebra of the form given by equation (\ref{qumator}) consistent Temperley-Lieb recoupling theory at a root of unity. This suggests that the delicate complex of issues related to quantum gravity with a cosmological constant and framed spin networks are not captured by the combinatorics of the Temperley-Lieb algebra at a root of unity, at least not as envisioned in the original work on $q$-quantum gravity \cite{qqg,qqgop,dis}.  While removing the assumption that $q$ is a root of unity might yield a consistent system, this would remove one of the nice features of the theory, that the cosmological constant provides a cutoff on the allowed representations.  However without a better understanding of the nature of framing, it is not clear which generalization is the best.  One promising direction of work to clear up the issue of framing is an investigation of the loop representation for $U(1)$ Chern-Simons theory in the context of the fractional quantum Hall effect \cite{dis}.

\begin{ack}
I thank Rouman Borrisov,
Robert Redfield, Rob Silversmith, and Lee Smolin for discussions.    This work was supported, in part, by Research Corporation.

\end{ack}

\appendix
\section{Recoupling theory}

I have have collected in this appendix all the relevant notation of
 Temperley-Lieb recoupling theory using the conventions of Kauffman and Lins \cite{KL}.
The complex phase $q$ is given by
$$
q =e^{i \pi / r}.
$$
for integer $r$. In, $q$-quantum gravity, the parameter $r$ is inversely related to the dimensionless cosmological constant $r \propto 1/\Lambda \, \ell_p^2 $.  For details on this relationship see \cite{dis} and \cite{leerev}.
The $q$-classical limit is when $q=1$ and $r \rightarrow \infty$ so that
$\hbar$ and/or $\Lambda$ tend to zero.
  
The basic irreducible representation is diagrammatically represented as a single line or ``strand.'' Closing
this line (or equivalently taking its trace) gives the loop value
$$
\kd{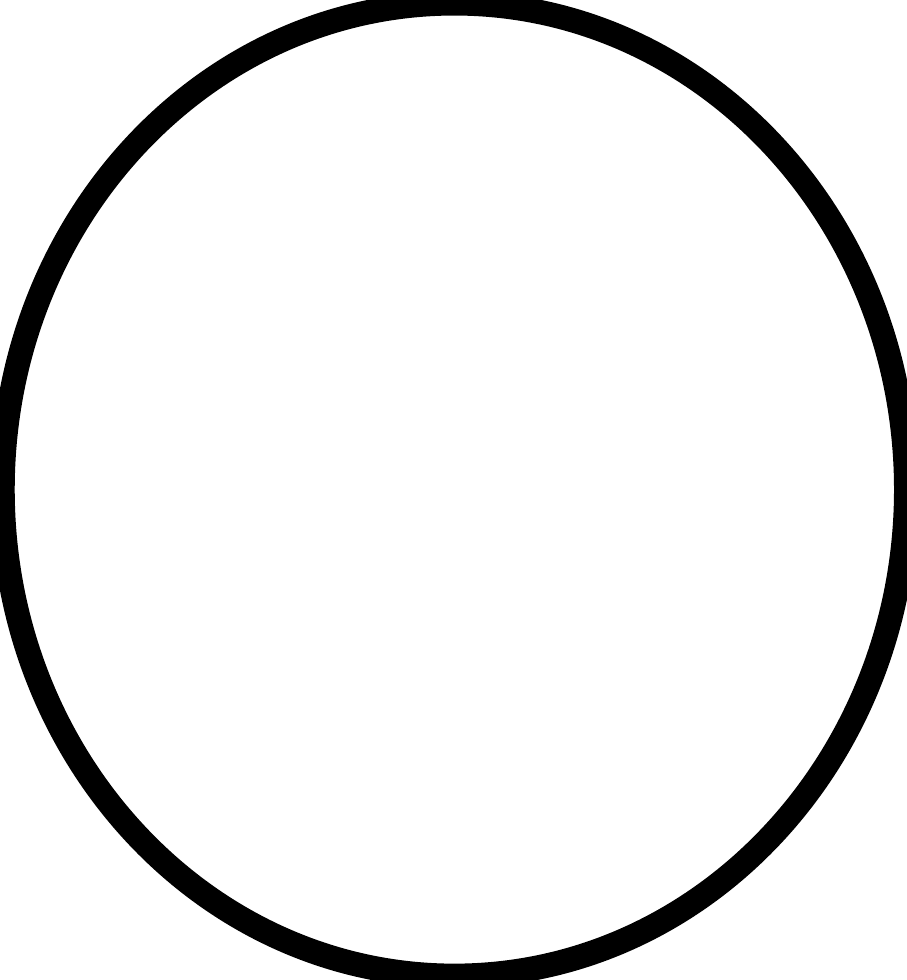} = -q - q^{-1} = \Delta_1.
$$
Higher representations   may be built from the basic line
using the Wenzel-Jones projector defined by
\begin{equation}
\label{WJP}
\kd{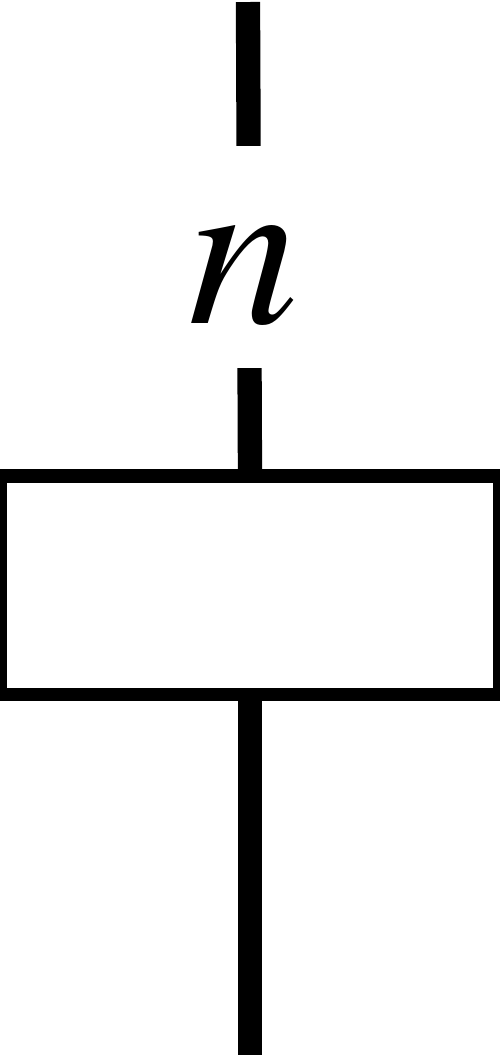} = \frac{ 1} { \{ n\}!} \sum_{\sigma \in S_n} \left( q^{-3/2} 
\right)^{|\sigma|} \kd{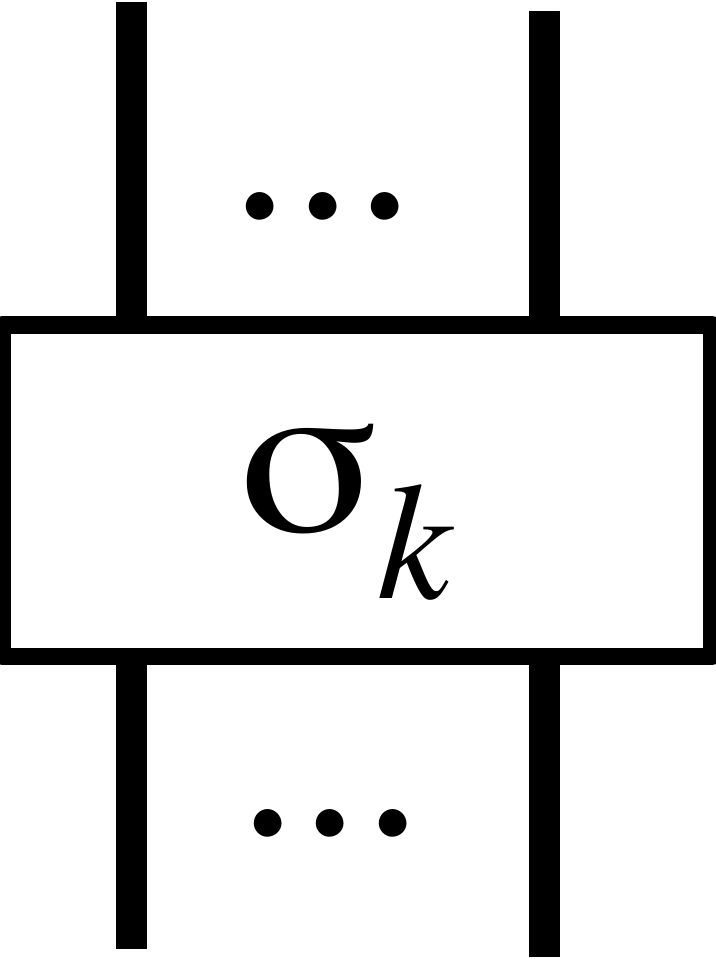}
\end{equation}
in which the sum is over elements of the symmetric group, $\sigma$;
$|\sigma|$ is the sign of permutation; the expansion $\kd{sgline}$ is
given in terms of the positive braid (the strands are only over crossed
$\kd{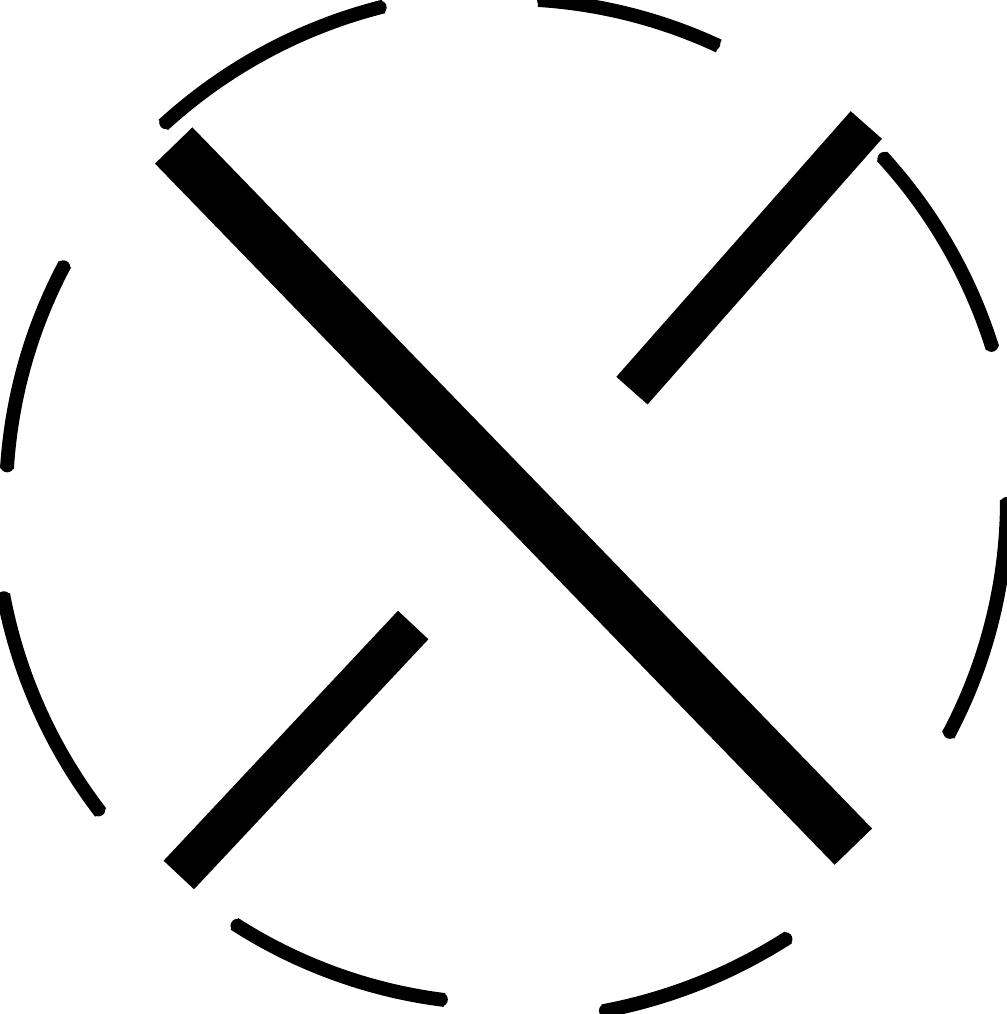}$); and the asymmetric quantum number ${n}$ is defined
by
\begin{equation}
\label{ASQN}
\{n \} := { 1 - q^{-2n} \over 1 - q^{-2}}.
\end{equation}
The quantum factorial is defined in the usual way $\{ n \} ! = \{ n \}
\{ n-1 \} \dots \{1\}$.  The Wenzel-Jones projector is irreducible since
$
\kd{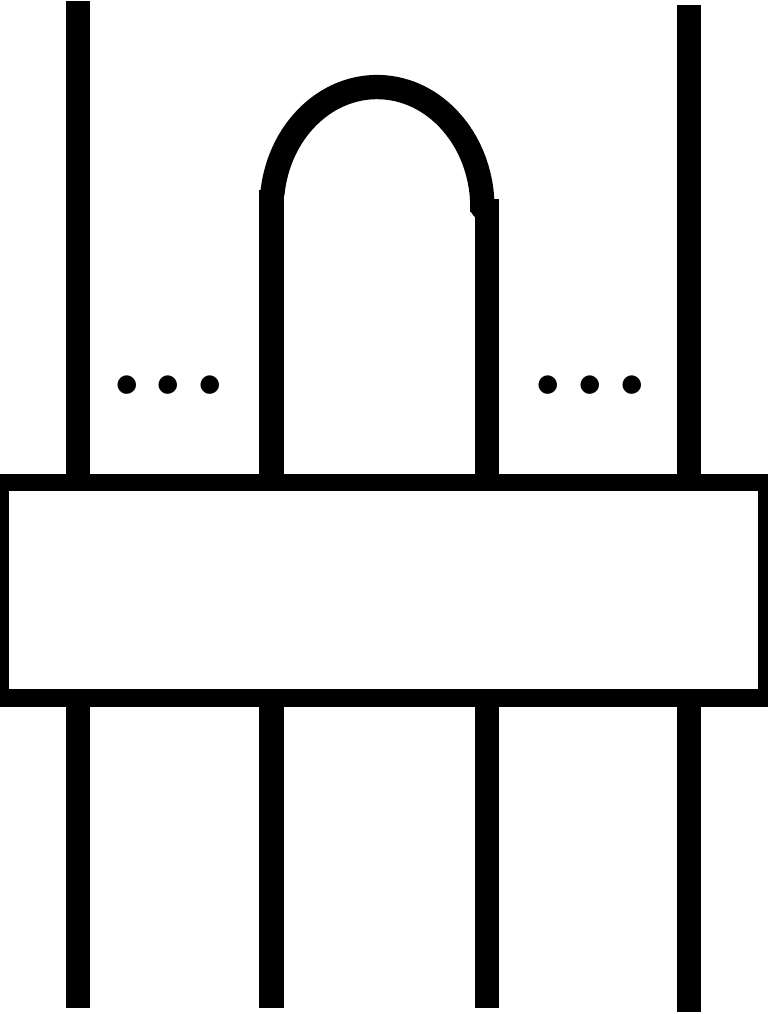} = 0.
$
There are finitely many representation as the projector vanishes for the $(r-1)$-th representation,
$
\kd{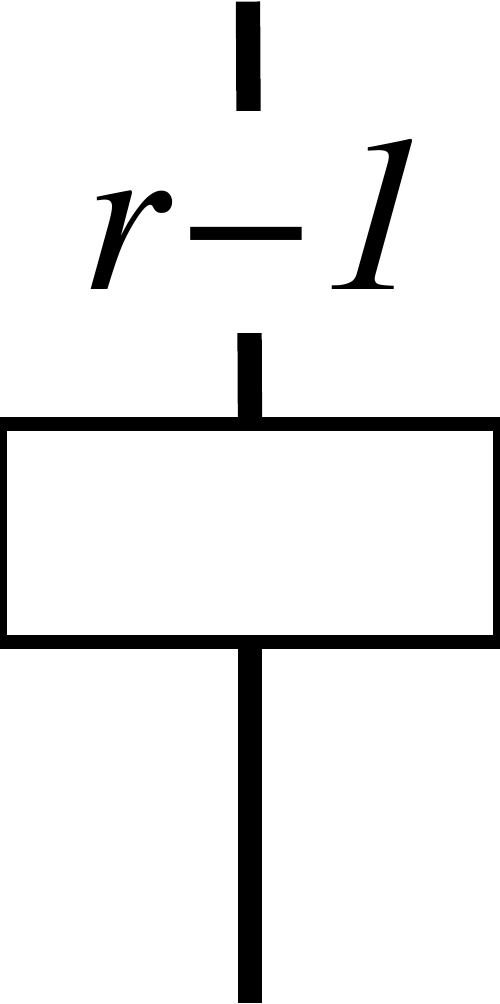} = 0.
$

The evaluation of a single
un-knotted  loop in the $n$ representation is 
$\Delta_n = (-1)^n [n+1]$
where $[n+1]$ is the dimension of the representation. This (symmetric) quantum number $[n]$ is given by
$$
[n] := { q^{n} - q^{-n} \over q - q^{-1} }.
$$

Representations may be added recursively using
the edge addition formula.  If a single path is labeled by $n$ and represented by $\ket{n}$ then
\begin{equation}
\label{eaf}
\ket{1} \ket{n} = \sum_{\epsilon = \pm 1} A(\epsilon) \ket{n+\epsilon}
\end{equation}
where $A(1) =1$ and $A(-1) = - [n]/[n+1]$.
Intertwiners are built from the basic trivalent vertex. The basic recoupling relation is
\begin{equation}
\kd{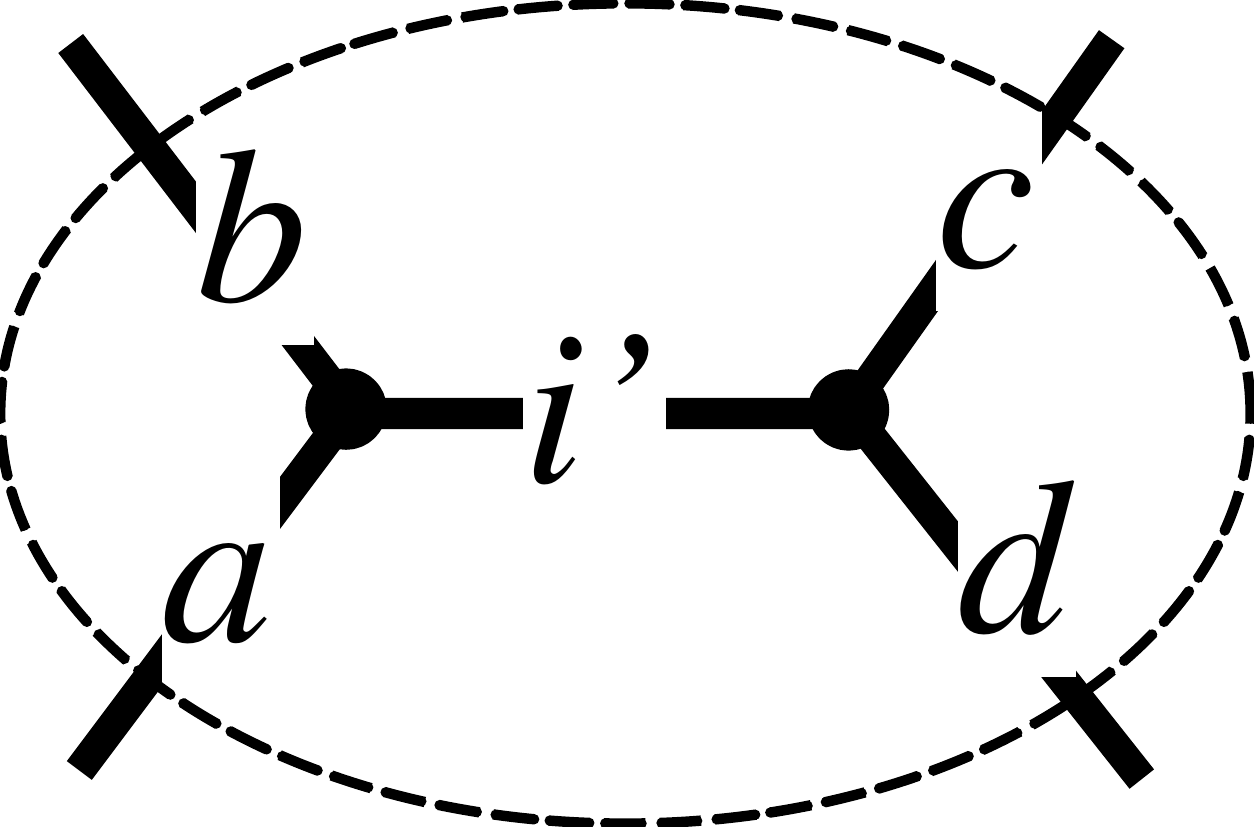} = \sum_{|a-b| \leq i \leq(a+b)}
\left\{ \begin{array}{ccc} a & b & i' \\ c & d & i 
\end{array} \right\} \kd{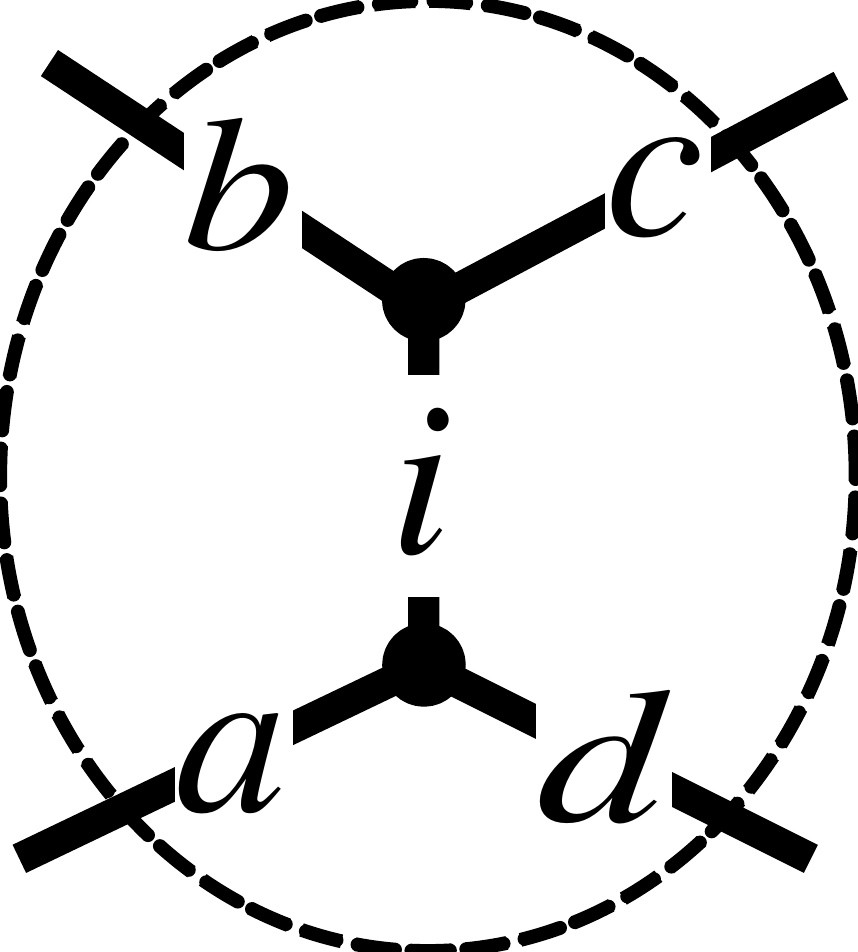}
\end{equation}
where on the right hand side is the $q-6j$-symbol is defined below. (The dashed lines emphasize that the recoupling is done only done at a point on a spin network.)

By Shur's lemma  the ``bubble'' diagram is proportional to a
single edge. A computation gives
\begin{equation} \label{bubble}
\kd{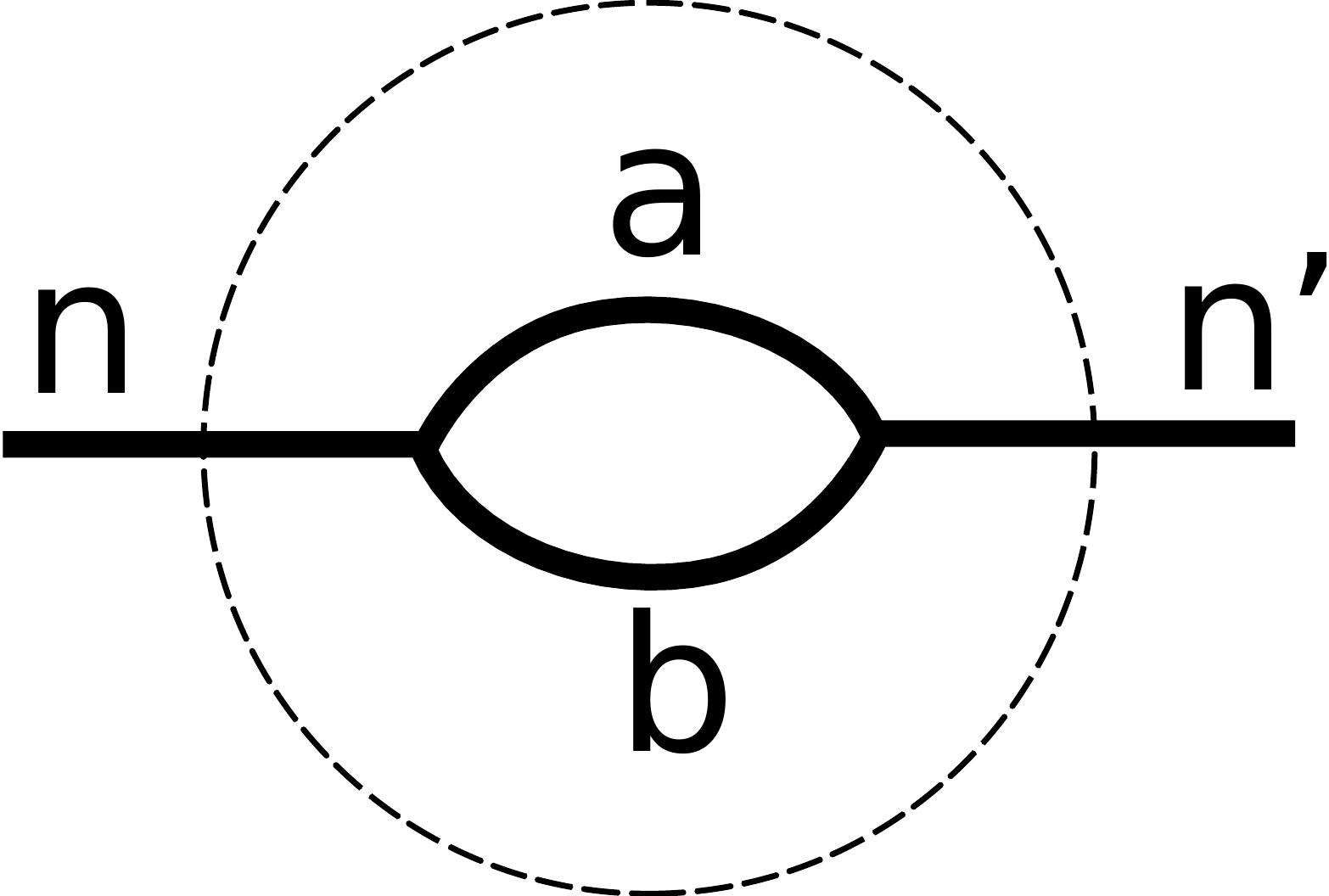} = \delta_{nn'}{ (-1)^{n} \theta(a, b, n) \over [n+1] }
\kd{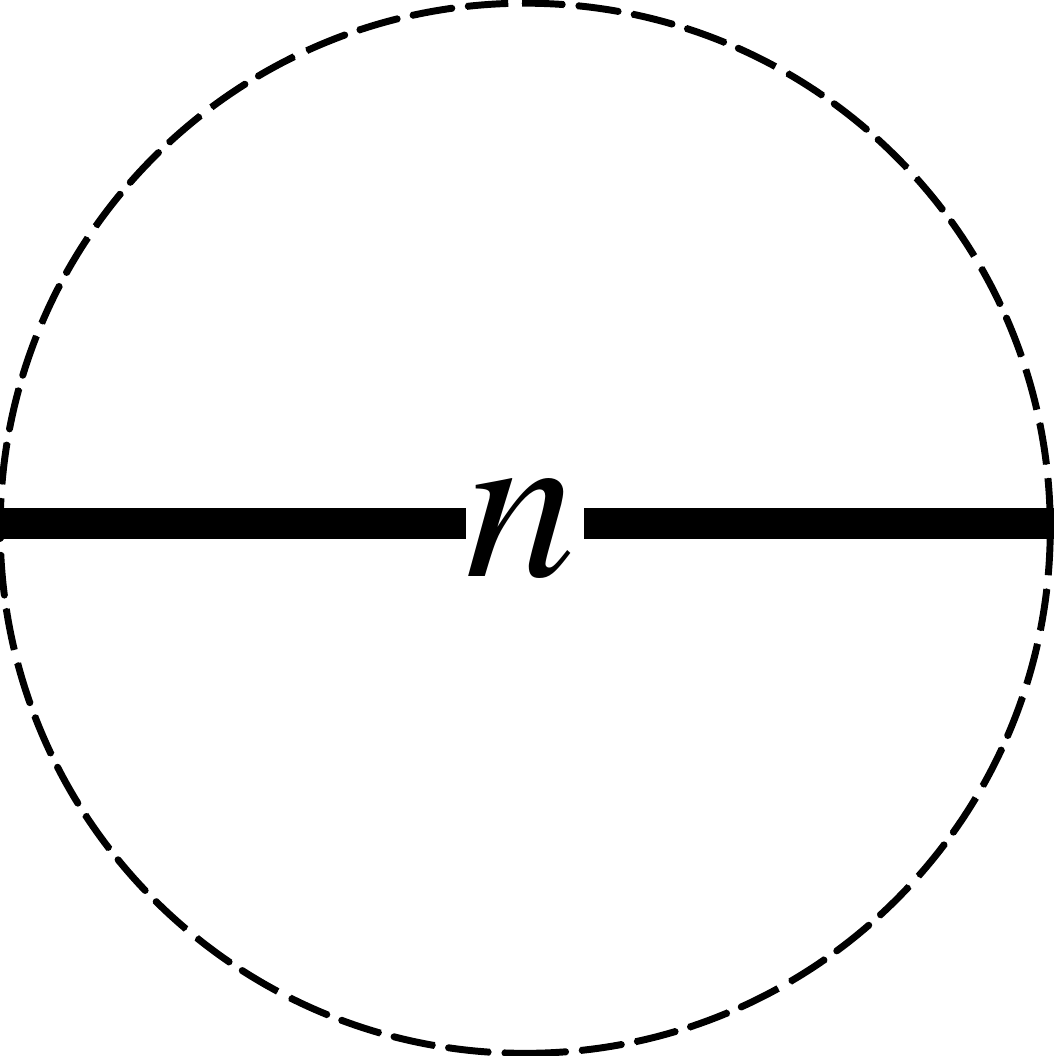}
\end{equation}
in which the function $\theta(a, b, n)$ is given by
\begin{equation} \label{theta}
\theta(m,n,l)= \kd{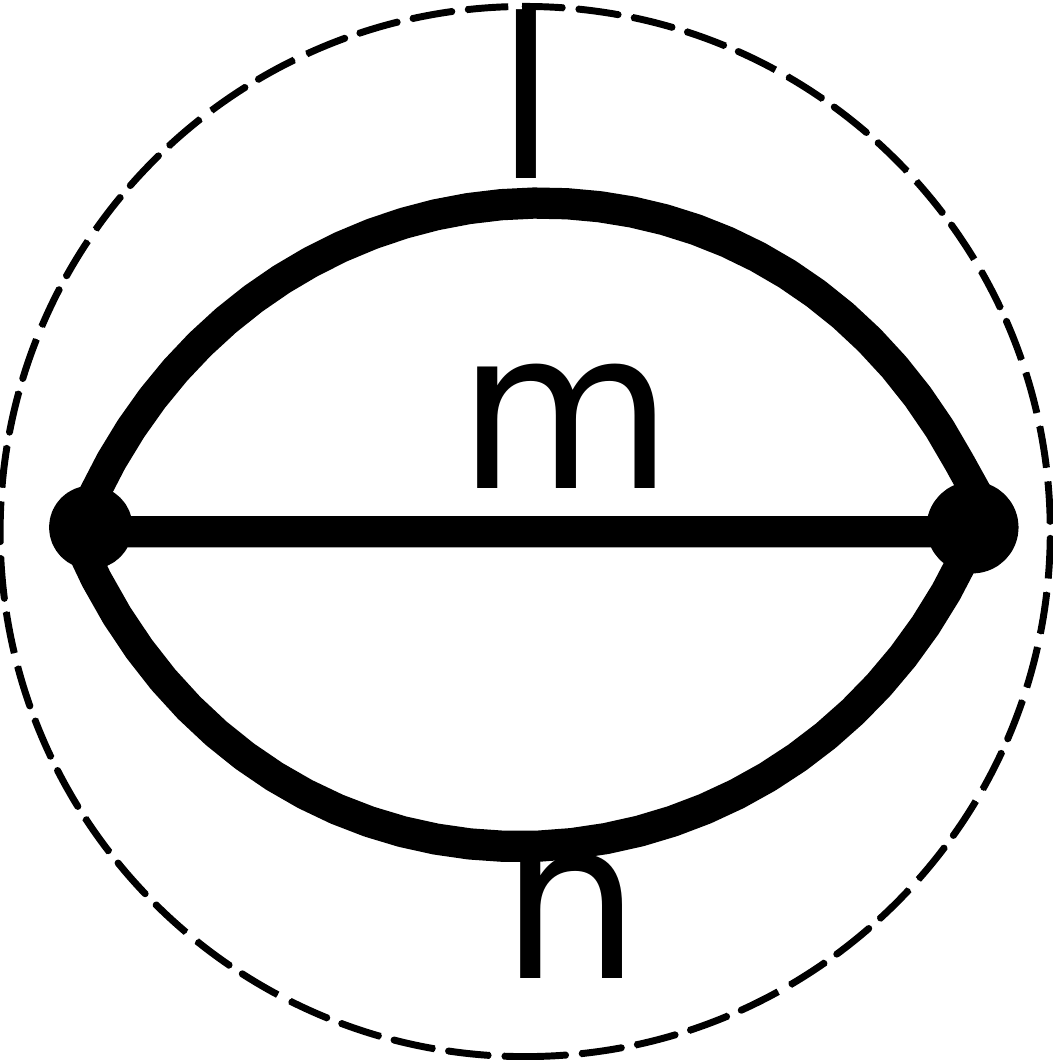} =
(-1)^{(a+b+c)}{[a+b+c+1]![a]![b]![c]! \over [a+b]![b+c]!
[a+c]!}
\end{equation}
where $a=(l+m-n)/2$, $b=(m+n-l)/2$, and $c=(n+l-m)/2$.

To evaluate diagrams required in the evaluation of the action of the algebra,
\begin{equation}
\label{diag}
\lkd{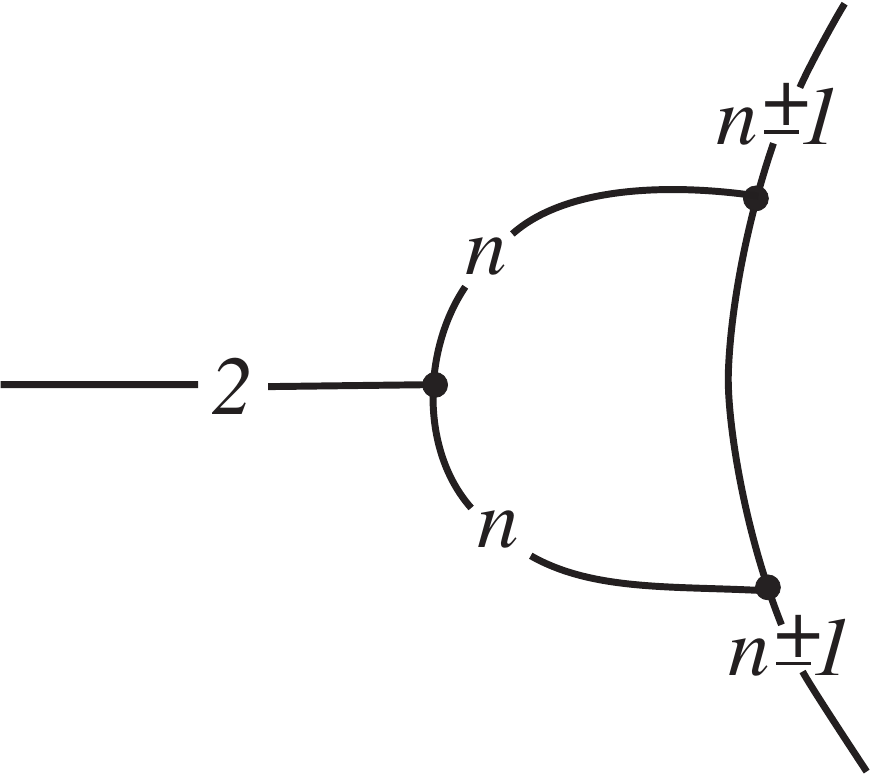},
\end{equation}
one needs the $q6j$-symbol or a $\text{Tet}$.  It is defined by
\begin{equation}
\begin{split}
\label{TetDef}
\kd{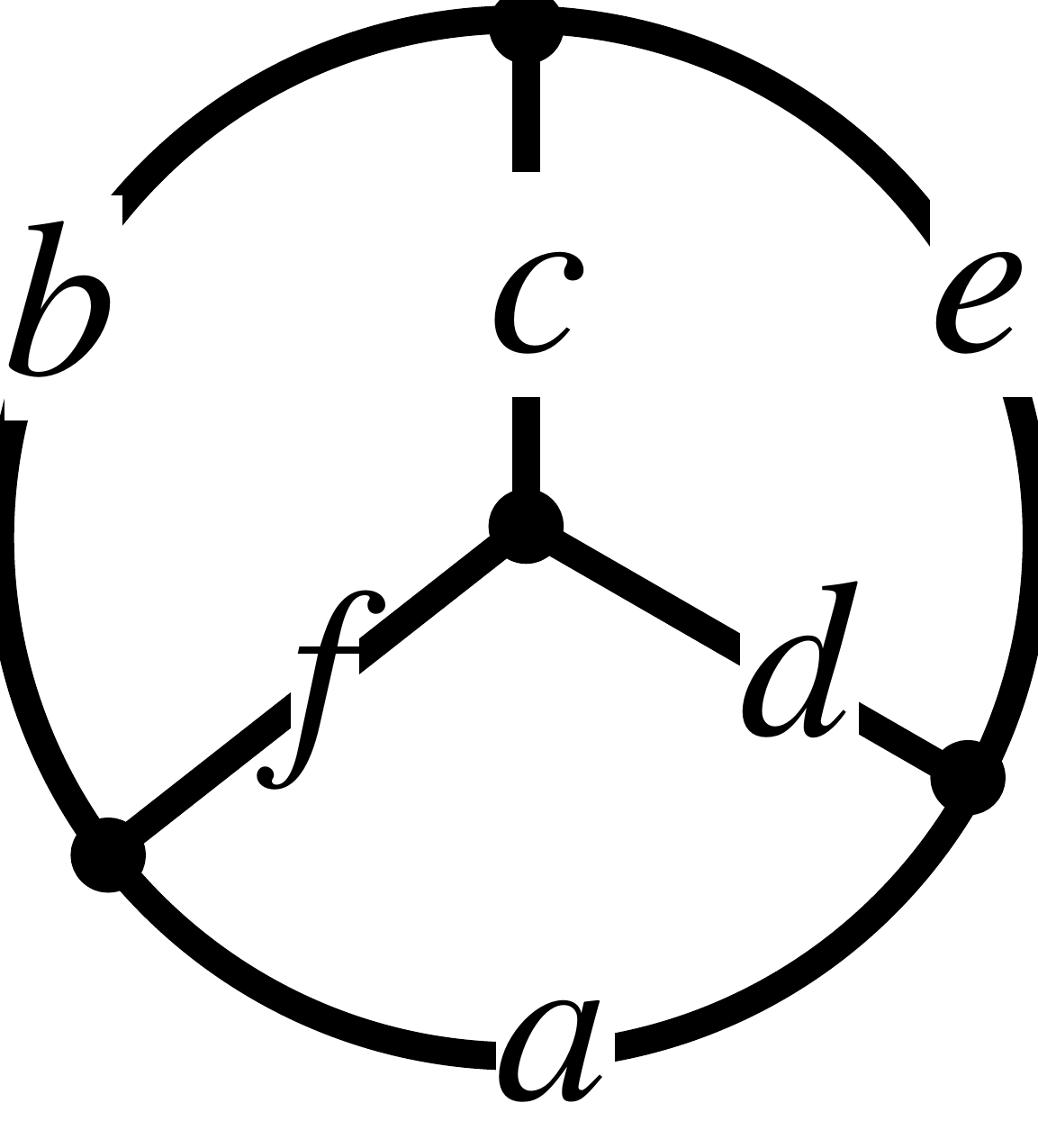} = \kd{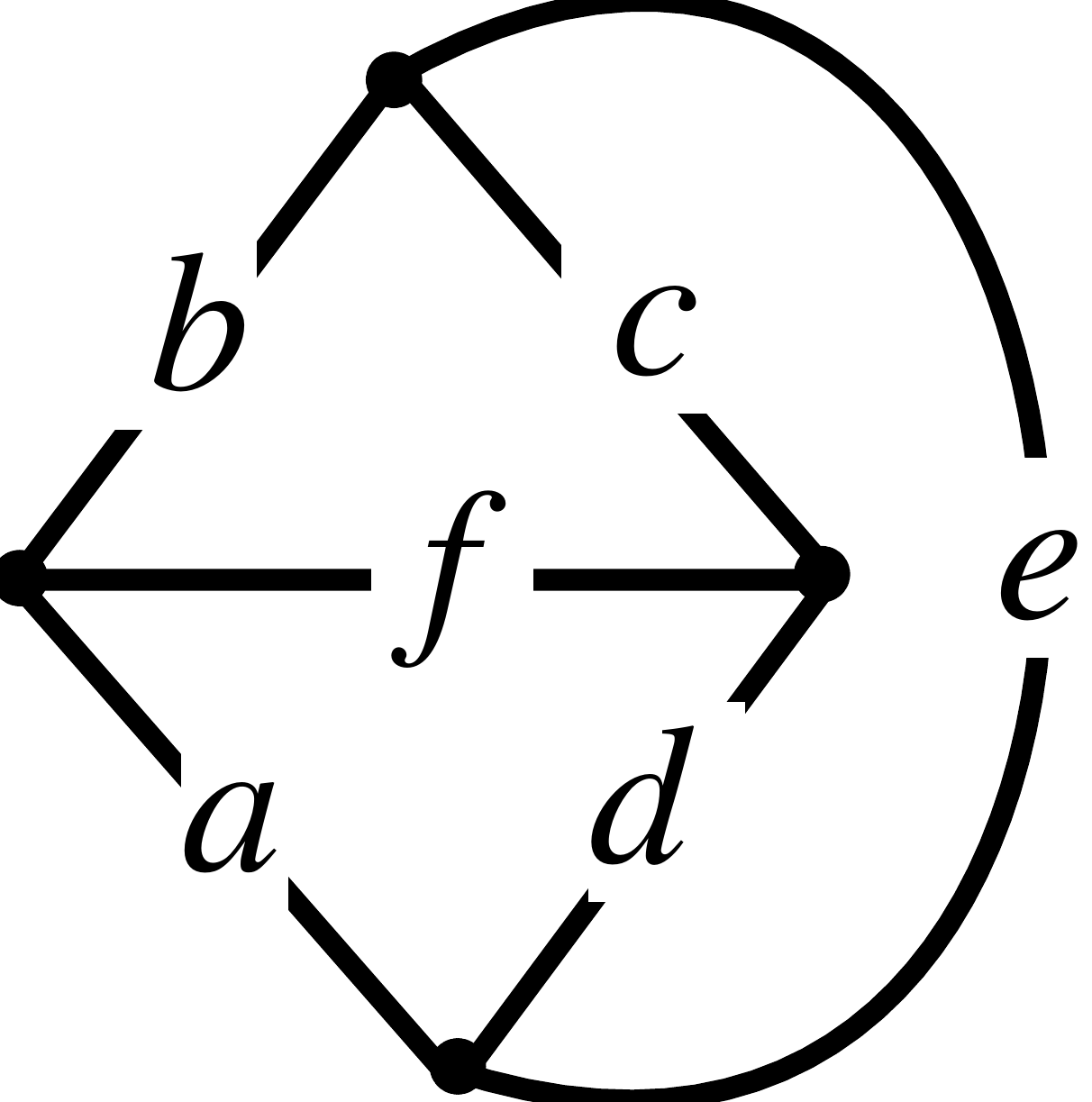} &\equiv \text{Tet} \begin{bmatrix} a & b & e \\ c & d & f \end{bmatrix} \\
\text{Tet} \begin{bmatrix} a & b & e \\ c & d & f \end{bmatrix} &= N 
\sum_{m \leq s \leq M} (-1)^s  {  [s+1]! \over
\prod_i\, [s-a_i]! \; \prod_j \, [b_j -s]! } \\
N &= { \prod_{i,j}\, [b_j - a_i]! \over [a]![b]![c]![d]![e]![f]!}
\end{split} \end{equation}
in which 
\begin{align*}
a_1 &= \tfrac{1}{2} ( a +d + e) & b_1 &= \tfrac{1}{2} ( b +d + e+ f) \\
a_2 &= \tfrac{1}{2} ( b +c + e) & b_2 &= \tfrac{1}{2} ( a +c + e +f) \\
a_3 &= \tfrac{1}{2} ( a +b + f) & b_3 &= \tfrac{1}{2} ( a +b + c+d) \\
a_4 &= \tfrac{1}{2} ( c +d + f) \\ 
\text{with } m &={\rm max}\, \{a_i\}  & M &={\rm min}\, \{b_j\}
\end{align*}

The 
$q-6j$ symbol is defined as
\begin{equation}
\label{q6jsym}
\left\{ \begin{array}{ccc} a & b & i \\ c & d & j \end{array} \right\}
:=
{ \text{Tet} \begin{bmatrix} a & b & i \\ c & d & j \end{bmatrix} \Delta_i
\over \theta(a,d,i) \; \theta(b,c,j) }
\end{equation}
It satisfies the Biedenharn-Elliot identity
\begin{equation}
\label{BE}
\sum_{i=0}^{r-1} 
\left\{ \begin{array}{ccc} a & b & i \\ c & d & j \end{array} \right\}
\left\{ \begin{array}{ccc} e & f & g \\ c & i & b \end{array} \right\}
\left\{ \begin{array}{ccc} e & g & h \\ d & a & i \end{array} \right\}
=
\left\{ \begin{array}{ccc} e & f & h \\ j & a & b \end{array} \right\}
\left\{ \begin{array}{ccc} h & f & g \\ c & d & j \end{array} \right\}
\end{equation}

The quantity that one needs to evaluate diagrams is the coefficient of a twist
\begin{equation} \begin{split} \label{lmove}
\kd{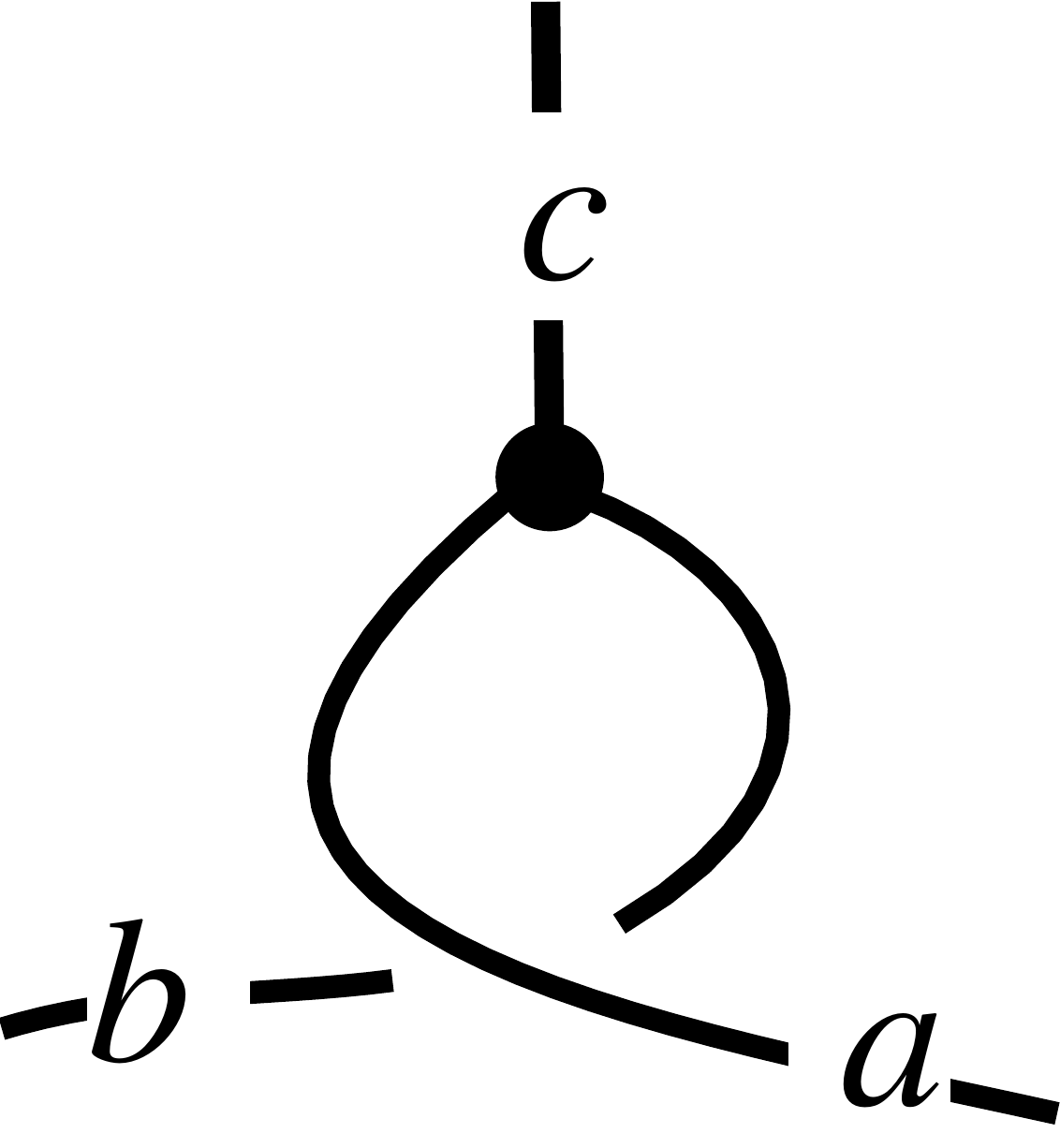} &= \lambda^{ab}_c \kd{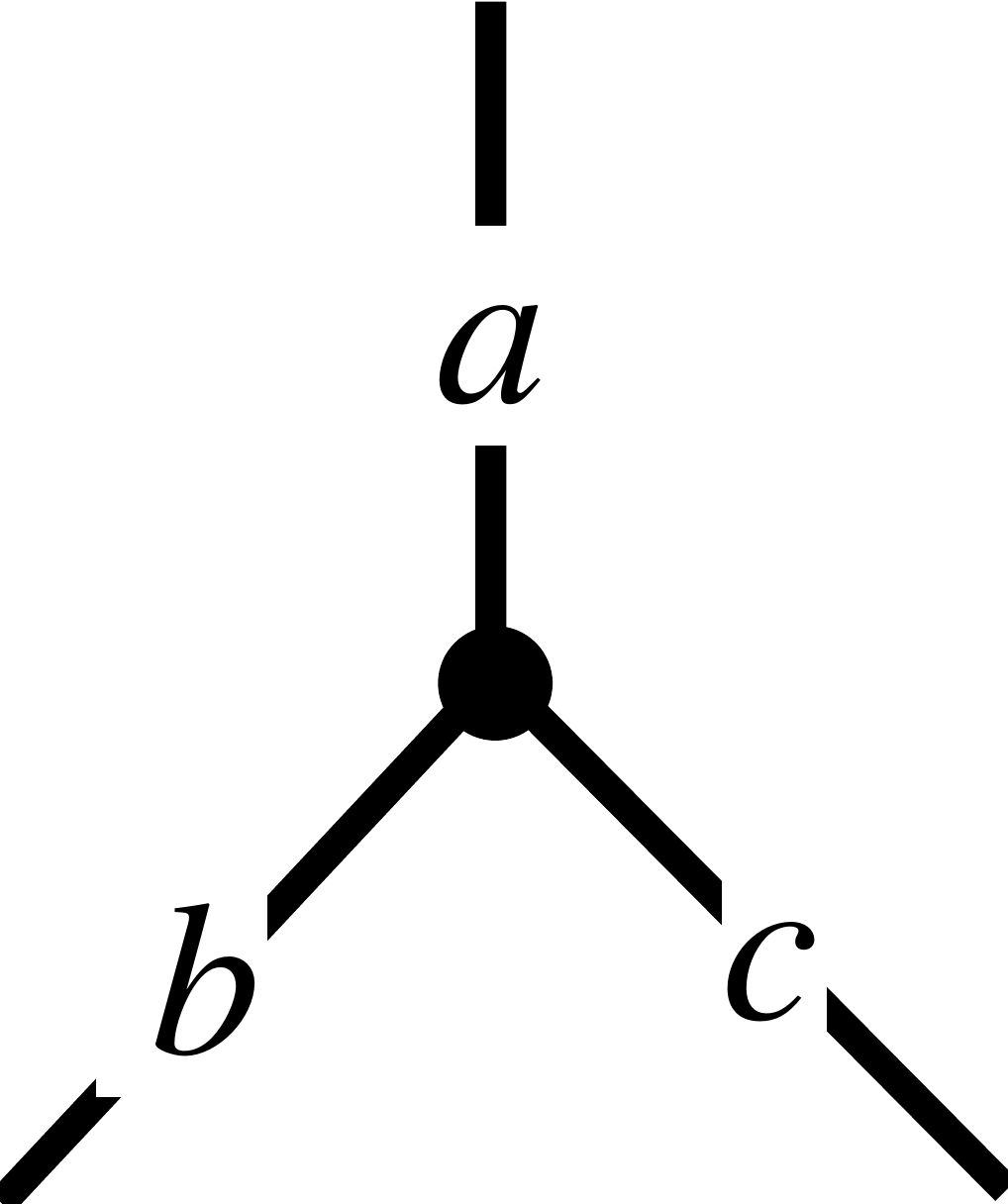}  \text{where $\lambda^{ab}_c$ is} \\
\lambda^{ab}_c &= (-1)^{(a+b-c)/2} A^{[a(a+2) + b(b+2) - c(c+2)]/2}
\end{split}
\end{equation}
For example, one has
$$
\kd{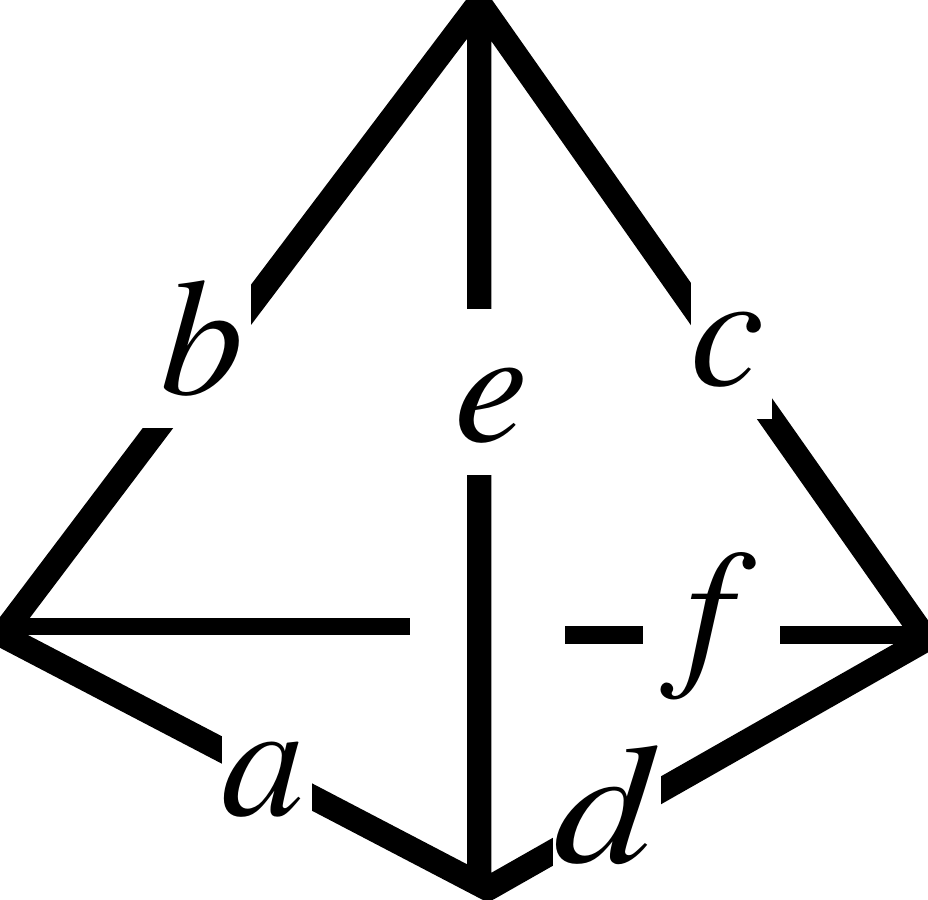} = \kd{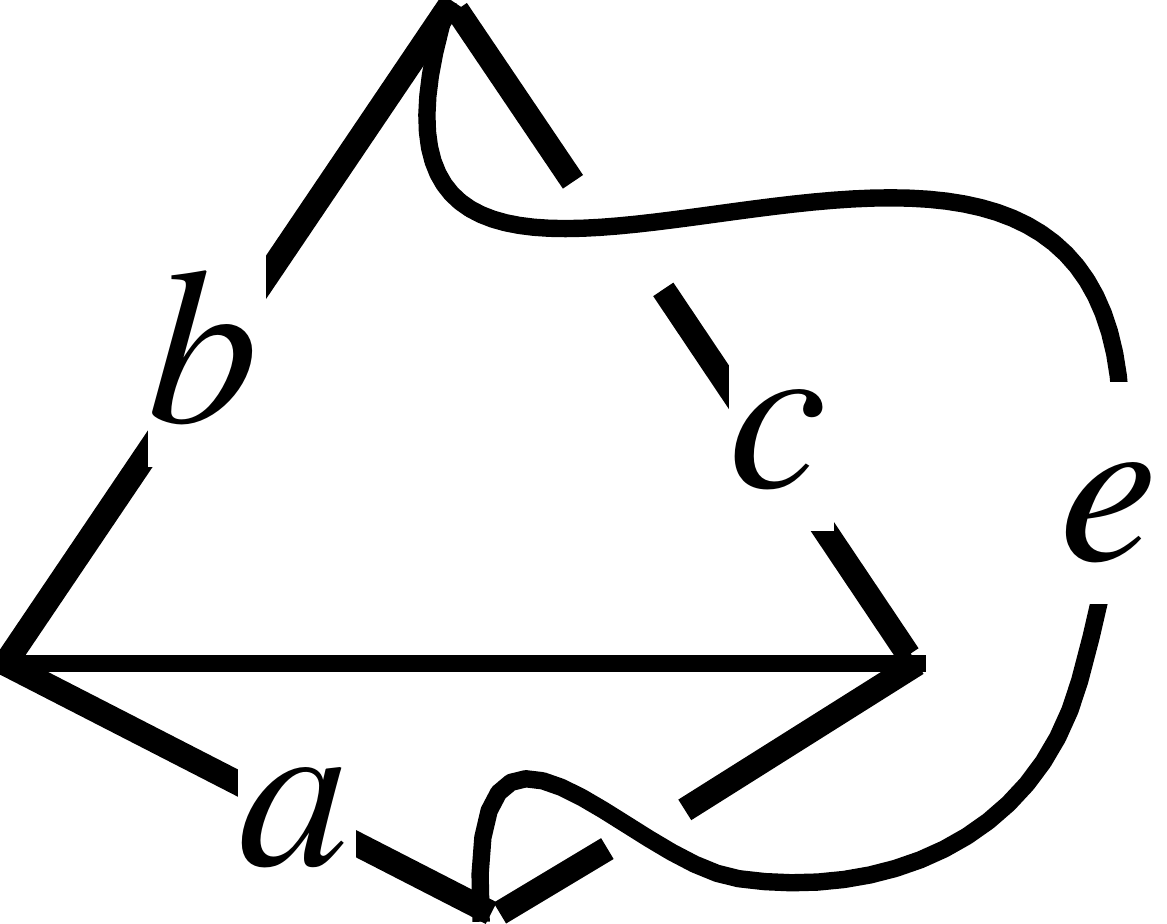} = \lambda^{ce}_b \; (\lambda^{be}_a)^{-1}
\kd{TetNet}.
$$ 

To evaluate the algebra one chooses a diagrammatic representation for the intersection, as in (\ref{diag}). Due to the regular isotopy of the diagrams the result does not depend on this choice. Following the conventions of \cite{KL} I do not include the normalization of the $q$-spin network state (a ``$1/\sqrt{\theta}$" for each 3-vertex). The first term of the algebra (\ref{alg}) follows immediately from the definitions giving $t_{n+1} \ket{ n+1, \, 1} + t_{n-1} \ket{n-1, \, 1} $. The second requires application of the edge addition formula and recoupling for the diagram
\begin{equation}
\label{algreduc1}
\lkd{alg2} = { \text{Tet} \begin{bmatrix} n \pm 1 & n \pm 1 & 1 \\ n & n & 2 \end{bmatrix}
\over \theta(n \pm 1,n \pm 1,2) } \lkd{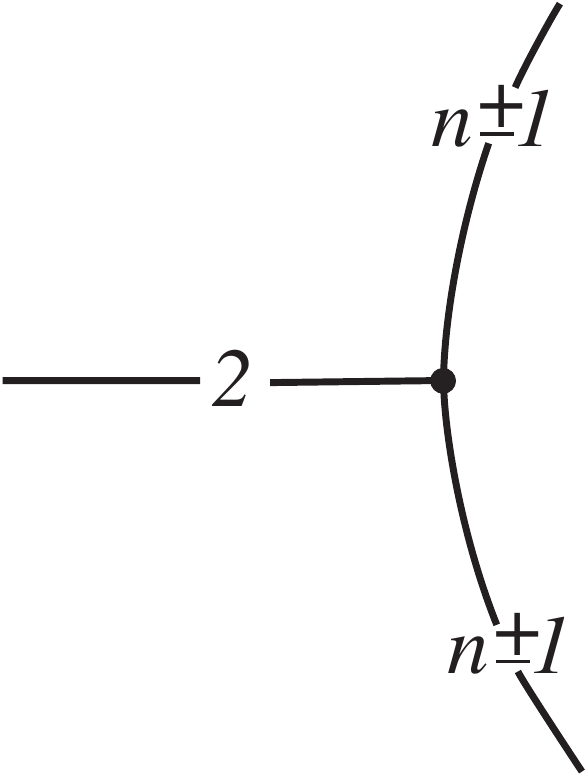}.
\end{equation}
Hence,
\begin{align}
S_\alpha T \ket{ n,  0} &= t_n \left( - \frac{[n]}{[n+1]} 
\frac{ \text{Tet} \begin{bmatrix} n - 1 & n - 1 & 1 \\ n & n & 2 \end{bmatrix}}{\theta(n - 1,n -1,2) } 
\ket{ n-1, 1}  + 
\frac{ \text{Tet} \begin{bmatrix} n + 1 & n + 1 & 1 \\ n & n & 2 \end{bmatrix}}{\theta(n + 1,n + 1,2) } 
\ket{n+1, 1} \right) \\
&= t_n \left( \frac{[n+2] [n-1]}{[n+1][n]} \ket{n-1, \, 1} + \ket{n+1, \, 1} \right).
\end{align}
Likewise the loop operator $S_{\alpha \beta}$ gives
\begin{align}
\label{algreduc2}
S_{\alpha \beta} \ket{n, \, 0}  &= - \frac{[n]}{[n+1]} 
\frac{ \text{Tet} \begin{bmatrix} n - 1 & n - 1 & n \\ 1 & 1 & 2 \end{bmatrix}}{\theta(n - 1,n - 1,2) } 
\ket{n-1, \, 1} + 
\frac{ \text{Tet} \begin{bmatrix} n + 1 & n + 1 & n \\ 1 & 1 & 2 \end{bmatrix}}{\theta(n + 1,n + 1,2) } 
\ket{n+1, \, 1} \\
&= - \frac{[n-1]}{[n+1]} \ket{n-1, \, 0} + \ket{n+1, \, 0}   
\end{align}

The general definition of the combinatoric operators $T$ and $S$ is determined through edge addition, grasping and recoupling.  For an arbitrary state $\ket{ m, \, n; \, i}$ for loops $\alpha$ and $\beta$ in a spin network and intertwiner $i$ (all other spin network labels are suppressed), $T$ acts by adding a edge addition and grasping while $S$ acts by edge addition.  Hence, $T$ acting on the state $\ket{m, \, n; \, i}$\footnote{I have chosen the intertwiner $i$ with vertices $(m,m,i)$ and $(n,n,i)$. The result does not depend on this choice.} gives, with the Biedenharn-Elliot identity (\ref{BE}),
\begin{equation}
\begin{split}
\label{tdef}
T \ket{m,n;i} = \sum_{k} \sum_{\epsilon_{1,2} =\pm 1} 
\frac{A(\epsilon_1) A(\epsilon_2)}{\Delta_{n+\epsilon_1}}
\left\{ \begin{array}{ccc} m & 2 & k \\ i & m & m \end{array} \right\}
\left\{ \begin{array}{ccc} k & 2 & n+\epsilon_1 \\ n+\epsilon_2 & n+\epsilon_1 & i \end{array} \right\}
\\
\times \text{Tet} \begin{bmatrix} n + \epsilon_1 & n + \epsilon_2 & n \\ 1 & 1 & 2 \end{bmatrix}  
\ket{m+\epsilon_1, \, n +\epsilon_2; \, k}
\end{split}
\end{equation}  
where the first sum is over admissible $k$'s. The definition of the operator $S_\alpha$ extends trivially but $S_{\alpha \beta}$ is more complex.  One way to write the result is
\begin{equation}
\label{sdef}
S_{\alpha \beta} \ket{m, \, n; \, i} = \sum_{\begin{aligned} \{\epsilon_j\}&=\pm 1\\ j&=1,2,3,4 \end{aligned}} B(m,n,i,\epsilon_j) \ket{m+\epsilon_1, \, n+\epsilon_2; \, i+\epsilon_3+ \epsilon_4} - C(m,n,i,\epsilon_j) \ket{m+\epsilon_1, \, n+\epsilon_2; \, i}
\end{equation}
with
\begin{equation} 
\begin{split}
B (m,n,i,\epsilon_j)= \prod_{j=1}^{4} A(\epsilon_j) 
\frac{ \text{Tet} \begin{bmatrix} m & m + \epsilon_1 & m \\ i+\epsilon_3 & i & 1 \end{bmatrix}}
{\theta(m + \epsilon_1 ,i+\epsilon_3,m) } 
\frac{ \text{Tet} \begin{bmatrix} n  & n + \epsilon_2 & n \\ i+\epsilon_3 & i & 1 \end{bmatrix}}
{\theta(n + \epsilon_2,i+\epsilon_3,n) }
\\ 
\times \frac{ \text{Tet} \begin{bmatrix} m + \epsilon_1 & m & m + \epsilon_1 \\ i+\epsilon_3 & i+\epsilon_3 +\epsilon_4 & 1 \end{bmatrix}}
{\theta(m + \epsilon_1,m + \epsilon_1,i+\epsilon_3 +\epsilon_4) } 
\frac{ \text{Tet} \begin{bmatrix} n + \epsilon_2 & n  & n + \epsilon_2 \\ i+\epsilon_3  & i+\epsilon_3 +\epsilon_4 & 1 \end{bmatrix}}
{\theta(n + \epsilon_2,n + \epsilon_2,i+\epsilon_3 +\epsilon_4) } 
\end{split}
\end{equation}
and
\begin{equation}
C(m,n,i,\epsilon_j) = \frac{A(\epsilon_1) A(\epsilon_2)}{[2]}
\frac{ \text{Tet} \begin{bmatrix} m & m + \epsilon_1 & i \\ m+\epsilon_1 & m & 1 \end{bmatrix}}
{\theta(m + \epsilon_1 ,m+\epsilon_1,i) } 
\frac{ \text{Tet} \begin{bmatrix} n & n + \epsilon_2 & i \\ n+\epsilon_2 & n & 1 \end{bmatrix}}
{\theta(n + \epsilon_2 ,n+\epsilon_2,i) } 
\end{equation}
If the loops $\alpha$ (or $\beta$) is a product of, say, $p$ edges then the factors above will also include $p-1$ recoupling factors for each virtual and real edge in the loop.

\end{document}